\documentclass[a4paper,11pt]{article}
\usepackage{aaskaiid}
\usepackage{orcidlink}
\usepackage{siunitx}
\DeclareSIUnit{\jansky}{Jy}

\title{Semi-empirical Predictions for Ultra-deep Radio Counts of Star-forming Galaxies with the SKAO}
\ShortTitle{Semi-empirical Predictions for SKAO}

\author[1]{Giulietti Marika\orcidlink{0000-0002-1847-4496}}
\ShortName{Giulietti et al.}  
\author[1]{Prandoni Isabella\orcidlink{0000-0001-9680-7092}}
\author[2]{Bisigello Laura\orcidlink{0000-0003-0492-4924}}
\author[1]{Bondi Marco}
\author[3]{Massardi Marcella\orcidlink{0000-0002-0375-8330}}
\author[3]{Bonato Matteo\orcidlink{0000-0001-9139-2342}}
\author[4]{Lapi Andrea\orcidlink{0000-0002-4882-1735}}

\affiliation[1]{INAF - Istituto di Radioastronomia, Via Piero Gobetti 101, 40129 Bologna, Italy}
\emailAdd{m.giulietti@ira.inaf.it}

\affiliation[2]{INAF - Osservatorio Astronomico di Padova, Via dell’Osservatorio 5, 35122 Padova, Italy}
\affiliation[3]{INAF - Istituto di Radioastronomia - Italian ALMA Regional Centre, Via Gobetti 101, 40129 Bologna, Italy}
\affiliation[4]{Scuola Internazionale Superiore di Studi Avanzati, Via Bonomea 265, 34136 Trieste, Italy }

\abstract{Star-forming galaxies (SFGs) dominate the faint radio sky at flux densities below 0.1\,mJy. Identifying these systems through a multiwavelength approach is essential to tracing the cosmic history of star formation. 
Upcoming surveys with the Square Kilometre Array Observatory (SKAO) in its AA4 configuration for the Mid array will probe these faint populations, offering unprecedented insights into the star formation activity of galaxies across cosmic time.
Semi-empirical models, built on minimal assumptions and empirical galaxy relations, provide an efficient framework to study galaxy evolution using recent radio and optical/near-infrared (NIR) data. We developed \texttt{SEMPER} (Semi-EMPirical model for Extragalactic Radio emission) to predict the radio luminosity functions and number counts of SFGs. \texttt{SEMPER} combines redshift-dependent stellar mass functions from deep NIR surveys with empirical relations such as the galaxy main sequence and the IR/radio correlation, to characterise the radio properties of massive, high-redshift galaxies. The model shows excellent agreement with recent deep radio observations and naturally predicts a substantial population of massive, dust-obscured galaxies already in place at early epochs. In this chapter, we extend the \texttt{SEMPER} framework to SKA surveys by including an evolving starburst fraction and computing differential number counts at 1.4\,GHz for both lensed and unlensed SFGs.  Furthermore, we predict the cosmic star formation rate density (SFRD) traced by radio-emitting galaxies up to $z\approx10$. Our results show that SKA surveys will probe the faintest flux-density regimes, dominated by galaxies powered by star formation, and that <20 hours of SKA-Mid Band 2 observations will recover at least $\approx$20\% of the total SFRD predicted by \texttt{SEMPER}, including contributions from optically/NIR-dark systems up to $z\approx 6$.}

\begin{document}
\include{journal-names}
\setlength{\bibsep}{0.0pt} 
\maketitle

\section{Introduction}

The next generation of radio facilities, led by the Square Kilometre Array Observatory (SKAO), will revolutionise our understanding of galaxy formation and evolution through radio observations. With its unprecedented sensitivity, angular resolution, and survey speed, the SKA will enable the detection of star-forming galaxies (SFGs) out to the epoch of reionisation, providing a dust-unbiased view of the cosmic star formation history.

Over the past decade, deep and wide-area radio surveys conducted with facilities, including SKA precursors and pathfinders, such as the \textit{Karl G. Jansky} Very Large Array (JVLA), LOFAR, ASKAP, and MeerKAT (\citealt{Helfand15}; \citealt{Smolcic2017}; \citealt{Prandoni2018}; \citealt{Lacy2020PASP..132c5001L};\citealt{Algera2020ApJ...903..139A}; \citealt{Norris2021PASA...38...46N}; \citealt{Heywood2022}; \citealt{Shimwell2022}) have reached sensitivities deep enough to probe the faint radio population dominated by radio-quiet active galactic nuclei and SFGs (\citealt{Smolcic2008, Smolcic2017}; \citealt{Padovani2015, Padovani2016}; \citealt{Prandoni2018}; \citealt{Algera2020ApJ...903..139A}; \citealt{vanderVlugt2021ApJ...907....5V}). These surveys have demonstrated the potential of the radio band to complement optical and infrared studies of galaxy evolution, particularly in the most dust-obscured regimes (\citealt{Smolcic2017}; \citealt{Best2023}).

Radio emission in SFGs arises from synchrotron radiation, produced by relativistic electrons accelerated in supernova remnants, and from free–free emission in ionised HII regions. This makes radio luminosity a powerful tracer of the star formation rate (SFR), provided that the contribution from AGN is negligible (\citealt{Condon2002AJ....124..675C}; \citealt{Kennicutt2012AR}). The tight correlation between radio and far-infrared (FIR) luminosities, the so-called far-infrared/radio correlation (FIRRC; \citealt{Helou1985}; \citealt{Condon1992}; \citealt{Murphy2011ApJ...737...67M}; \citealt{Delvecchio2021}), further supports the use of radio emission as a dust-independent tracer of star formation activity, even in highly obscured environments.

SKA and its precursor facilities will take this progress to the next level. The planned continuum surveys (\citealt{Prandoni2015}), covering hundreds to thousands of square degrees down to sub–$\mu$Jy sensitivities, will provide statistically significant samples of SFGs up to redshift $z \gtrsim 10$. These datasets will offer a unique opportunity to test models linking galaxy stellar mass, star formation, and radio emission across cosmic time.

Semi-empirical models provide a powerful way to interpret such forthcoming observations, along with state-of-the-art results from recent facilities, such as \textit{Euclid}, and the James Webb Space Telescope (JWST) in the optical/NIR regime of the electromagnetic spectrum. These models connect key galaxy properties (e.g. stellar mass, SFR, radio luminosity) through empirically calibrated relations, enabling efficient, data-driven predictions while keeping the parameter space limited (\citealt{Moster2018MNRAS.477.1822M}; \citealt{Behroozi2019MNRAS.488.3143B}; \citealt{Lapi2025}).

In this work, we employ our Semi-EMPirical model for Extragalactic Radio emission (\texttt{SEMPER}; \citealt{Giulietti2025}) to generate updated predictions for SKAO observations. The model has been successfully tested on radio data at 1.4 GHz and 150 MHz, showing remarkable agreement with the latest determinations from SKA pathfinders such as the Low-Frequency Array (LOFAR, \citealt{vanHaarlem2013}) and the Giant Metrewave Radio Telescope (GMRT). Moreover, \texttt{SEMPER} can reproduce the observed luminosity functions obtained for samples of radio-selected massive and dust-obscured SFGs at $z>3.5$, some of which contain a fraction of optical/NIR dark galaxies (see e.g. \citealt{Talia2021}; \citealt{Enia2022ApJ...927..204E}; \citealt{vanderVlugt2023ApJ...951..131V}), indirectly suggesting a natural link between this population and the radio-bright SFGs population observed at high redshift in deep radio fields. In recent years, several observational studies (e.g. \citealt{Wang2019}; \citealt{Gruppioni2020}) have highlighted that this population of dark galaxies can significantly impact the cosmic SFR density (SFRD), with a contribution up to $\approx 40 \%$ the contribution of high-redshift Lyman-Break Galaxies (\citealt{Talia2021}; \citealt{Enia2022ApJ...927..204E}; \citealt{Behiri2023}; \citealt{vanderVlugt2023ApJ...951..131V}; \citealt{Gentile2024a}).

This paper is organised as follows. Section \ref{sec:model} outlines \texttt{SEMPER} and its predictions for the SFRD. In Sect.\ref{sec:ska_pred}, we compare our updated model's number counts at 1.4 GHz with the depths expected for surveys conducted with the SKA-Mid AA4 configuration, along with the predicted SFRD computed from SKA observations.
Finally, we summarise our results in Sect. \ref{sec:summmary}.
We assume a \cite{Chabrier2003} initial mass function (IMF) and a standard $\Lambda$CDM cosmology with parameters: $H_0=70 \rm \, km\,s^{-1}\, Mpc^{-1}$, $\Omega_{\Lambda}=0.7$ and $\Omega_{\rm m}=0.3$, such that $h_{70} \equiv  H_0 /(70\, \rm km \, \rm s^{-1} \rm Mpc^{-1}) = 1$. The radio source spectra are assumed to be described by a simple power law $S_{\nu} \propto \nu^{\alpha}$, where $S_{\nu}$ is the monochromatic flux density at a certain frequency $\nu$ and $\alpha$ is the radio spectral index.

\section{The model}\label{sec:model}

In \cite{Giulietti2025}, we presented \texttt{SEMPER} and compared its predictions with the observed luminosity functions and number counts at 1.4\,GHz and 150\,MHz. In the following, we summarise the main ingredients of the model and the updates introduced in the present analysis.

\texttt{SEMPER} was built by combining up-to-date empirical relations, starting from the redshift-dependent galaxy stellar mass function (SMF), based on the recent deep NIR data from the COSMOS2020 catalogue (\citealt{Weaver2023}), with up-to-date observed scaling relations: the galaxy main sequence from \cite{Popesso2023} and the mass- and redshift-dependent IRRC from \cite{Delvecchio2021} and \cite{McCheyne2022}, drawn at 1.4 GHz and 150 MHz, respectively. 

We fitted the SMF for SFGs from the COSMOS2020 catalogue for the redshift range $0.2\leq z <5$ along with the local SMF ($z<0.08$) from \cite{Driver2022}. For this purpose, we adopted a double power-law profile, which accounts for the excess of galaxies at high stellar masses for $z\gtrsim$. This approach is analogous to what has been adopted by \cite{Shuntov2025} for SMF obtained from recent JWST observations. The Double Power-law has the form:

\begin{equation}\label{eq:double_powerlaw}
\begin{aligned}
\log \Phi d \log M_{\star}=- & \log \left(10^{\left(\log M_{\star}-\log M_0\right)(\alpha+1)+\log \Phi_1}\right.
+  \left.10^{\left(\log M_{\star}-\log M_0\right)(\beta+1)+\log \Phi_2}\right) d \log M_{\star},
\end{aligned}
\end{equation}

\noindent
where $\log \Phi_1$ and $\log \Phi_2$ are the normalisations of the two power laws, $\alpha$ and $\beta$ are the two slopes, $M_{\star}$ is the stellar mass, and $M_0$ represents the mass corresponding to the slope-change. To allow a full interpolation of our model over a broad redshift range, we fitted the best-fitting parameters of the double power-law profile as a function of redshift, enabling us to recover the shape and evolution of the SMFs also in redshift intervals not directly sampled by observations.

The SMFs were then convolved with the main sequence (\citealt{Brinchmann2004MNRAS.351.1151B}; \citealt{Noeske2007ApJ...660L..43N}) relation from \cite{Popesso2023}. 
In \cite{Giulietti2025}, we assumed SFGs at a fixed redshift and stellar mass to be distributed in SFR as a double Gaussian profile (\citealt{Bethermin2012}; \citealt{Sargent2012}; \citealt{Ilbert2015}; \citealt{Schreiber2015}). In this assumption, the two Gaussian distributions represent main sequence galaxies, which constitute the dominant population, and starburst galaxies, which are distributed approximately $3-4\sigma$ above the main sequence, and are described as:

\begin{equation}\label{eq:double_gaussian_SFR}
\begin{aligned}
\frac{d p}{d \log \psi}\left(\psi \mid z, M_{\star}\right)= & \left(\frac{A_{\mathrm{MS}}}{\sqrt{2\pi \sigma^2_{\rm MS}}}\right) \exp \left[-\frac{\left(\log \psi-\langle\log \psi\rangle_{\mathrm{MS}}\right)^2}{2 \sigma_{\mathrm{MS}}^2}\right] \\
& +\left(\frac{A_{\mathrm{SB}}}{\sqrt{2\pi \sigma^2_{\rm SB}}}\right) \exp \left[-\frac{\left(\log \psi-\langle\log \psi\rangle_{\mathrm{SB}}\right)^2}{2 \sigma_{\mathrm{SB}}^2}\right].
\end{aligned}
\end{equation}

In the above equation, $\psi$ is the SFR, $\langle  \log\psi \rangle_{\rm MS}$ represents the first Gaussian's central value and refers to the main sequence (MS), and $\langle  \log\psi \rangle_{\rm SB} = \langle  \log\psi \rangle_{\rm MS} + 0.59$ is the central value of the second Gaussian referring to starburst (SB) galaxies. A$_{\rm MS}$ and A$_{\rm SB}$ are the fractions of MS and starburst galaxies, respectively. The one-sigma dispersions of the first and second Gaussians are $\sigma_{\rm MS}=0.188$ and $\sigma_{\rm SB}=0.243$.
In \cite{Giulietti2025}, the starburst fraction A$_{\rm SB}$ was kept fixed at the value of 0.03, following \cite{Sargent2012}, assuming therefore this value to remain constant with redshift and stellar mass. In this work, we follow recent results indicating an increase in this quantity, compared to the MS, for low mass ($M_{\star} \leq 10^9$) or higher redshift ($z\geq 2- 3$) galaxies (e.g. \citealt{Caputi2017}; \citealt{Bisigello2018}; \citealt{Chruslinska2021MNRAS.508.4994C}; \citealt{Rinaldi2025}). For example, \cite{Bisigello2018} predict the starburst fraction to vary from $\approx 5\%$ at $z=0.5$--$1.0$ up to $\approx 16\%$ at $z=2.0$--$3.0$ with respect to the total number of galaxies at stellar masses of $\log(M_{\star}/M_{\odot}) = 8.25-11.25$, while for the low mass regime ($\log(M_{\star}/M_{\odot})<9$), the number of starburst galaxies varies from $\lesssim 20\%$ at $z\approx0.75$, and between $20$ and $30\%$ at $z\approx1.5$. Similar results were found in \cite{Rinaldi2025}, which extended the redshift range of up to $z\approx 7$, by exploiting JWST data. Therefore, we update our previous prescriptions by also adopting the evolving starburst fractions from \cite{Bisigello2018} and \cite{Rinaldi2025}. 

The result of the convolution between the SMF and the MS gives the star formation rate function (SFRF):

\begin{equation}\label{eq:sfrf}
\begin{aligned}
\frac{d^2 N_{\mathrm{SMF}+\mathrm{MS}}}{d \log \psi d V}(z, \log \psi)= & \int d \log M_{\star} \frac{d^2 N_{\rm SFG}}{d \log M_{\star} d V}\left(z, \log M_{\star}\right)
 \times \frac{d p}{d \log \psi}\left(\log \psi \mid z, M_{\star}\right),
\end{aligned}
\end{equation}

where $V$ is the comoving cosmological volume.

\subsection{Radio Luminosity Function and Number Counts}

The radio luminosity function (LF) for SFGs can be derived by expressing the SFRF in terms of the radio luminosity $L_{\nu}$ at a given frequency $\nu$. For this purpose, we adopted the stellar mass- and redshift-dependent $L_{\rm SFR}$--radio correlation from \cite{Delvecchio2021}, defined by the parameter $q_{\rm UV+IR}$:

\begin{equation}\label{eq:firrc_delvecchio}
\begin{aligned}
q_{\rm UV+IR}(M_{\star},z)=(2.743\pm0.034)\times A^{(-0.025\pm0.012)} -B\times(0.234\pm0.017).
   \end{aligned}
\end{equation}

Where $A=(1+z)$, $B=\log({M_{\star}/M_{\odot})}-10$. $q_{\rm UV+IR}=q_{\rm SFR}$ includes the contribution for the dust-uncorrected UV emission (see Sec. 2.3 of \citealt{Giulietti2025} for details) and is expressed as:

\begin{equation}\label{eq:firrc}
q_{\rm UV+IR}=\log \left(\frac{L_{\mathrm{SFR}}[\mathrm{W}] / 3.75 \times 10^{12}}{L_{1.4 \mathrm{GHz}}\left[\mathrm{W}\, \mathrm{Hz}^{-1}\right]}\right).
\end{equation}

In the above expression, $L_{\rm 1.4 GHz}$ is the rest-frame radio luminosity at 1.4\,GHz, and $L_{\rm SFR}$ is the luminosity obtained from $\psi$ using the relations from \cite{Kennicutt2012AR} rescaled to a \cite{Chabrier2003} IMF.

We then combined Eqs. \ref{eq:firrc_delvecchio}, \ref{eq:firrc}, and \ref{eq:gaussianprob} with Eq. \ref{eq:sfrf} and convolved the result with a Gaussian distribution representing the probability of having a given radio luminosity at the frequency $\nu$ ($L_{\nu}$), at fixed $\psi$, $M_{\star}$ and $z$:

\begin{equation}\label{eq:gaussianprob}
    \begin{aligned}
    \frac{d \rm p}{d \log{L_{\nu}}} \left(\log L_{\nu} \mid \psi, M_{\star},z \right) =  
        \left(\frac{1}{\sqrt{2\pi\sigma^2_{\rm q_{\rm UV+IR}}}} \right) \exp \left[ - \frac{\left( \log L_{\nu} - \langle \log L_{\nu} \rangle \right)^2}{2 \sigma^2_{\rm q_{\rm UV+IR}}} \right].
    \end{aligned}
\end{equation}

\noindent
The term $\sigma_{\rm UV+IR}$ accounts for the scatter of Eq. \ref{eq:firrc_delvecchio} and $\langle \log L_{\nu} \rangle$ is the radio luminosity corresponding to a given $\psi$, and it is derived from Eq. \ref{eq:firrc} by rescaling the frequency $\nu$ from 1.4 GHz. 
The final expression for the LF of SFGs, obtained by combining Eq. \ref{eq:sfrf}, with Eq. \ref{eq:firrc} and Eq. \ref{eq:gaussianprob}, is:

\begin{equation}\label{eq:final}
    \begin{aligned}
     \frac{d^2\rm N}{d\log L_{\nu}dV}\left(\log L_{\nu}, z\right) = \int  d\log M_{\star} \frac{d^2N_{\rm SFG}}{d\log M_{\star}dV} \left(\log M_{\star} \mid z \right) \\ \times \int d\log \psi \frac{dp}{d\log \psi} \left(\psi \mid z, M_{\star} \right)  \\ \times \frac{d \rm p}{d \log{L_{\nu}}} \left(\log L_{\nu} \mid \psi, M_{\star},z \right).
    \end{aligned}
\end{equation}

By integrating the above equation over the redshift, one obtains:

\begin{equation}\label{eq:number_counts}
\frac{d^2 N}{d \log S_\nu d \Omega}\left(S_\nu\right)=\int d z \frac{d^2 V}{d z d \Omega} \frac{d^2 N}{d \log L_\nu d V}\left(L_{\nu(1+z)}, z\right),
\end{equation}

\noindent
which is the expression for the radio-band number counts. $dV/dz d\Omega$ is the cosmological volume per unit solid angle, and $S_{\nu}$ is the observed flux defined as:

\begin{equation}
S_\nu=\frac{L_{\nu(1+z)}(1+z)}{4 \pi D_L^2(z)},
\end{equation}

\noindent
with $D_L(z)$ being the luminosity distance.

\begin{figure}{h}
    \centering
    \includegraphics[width=0.7\linewidth]{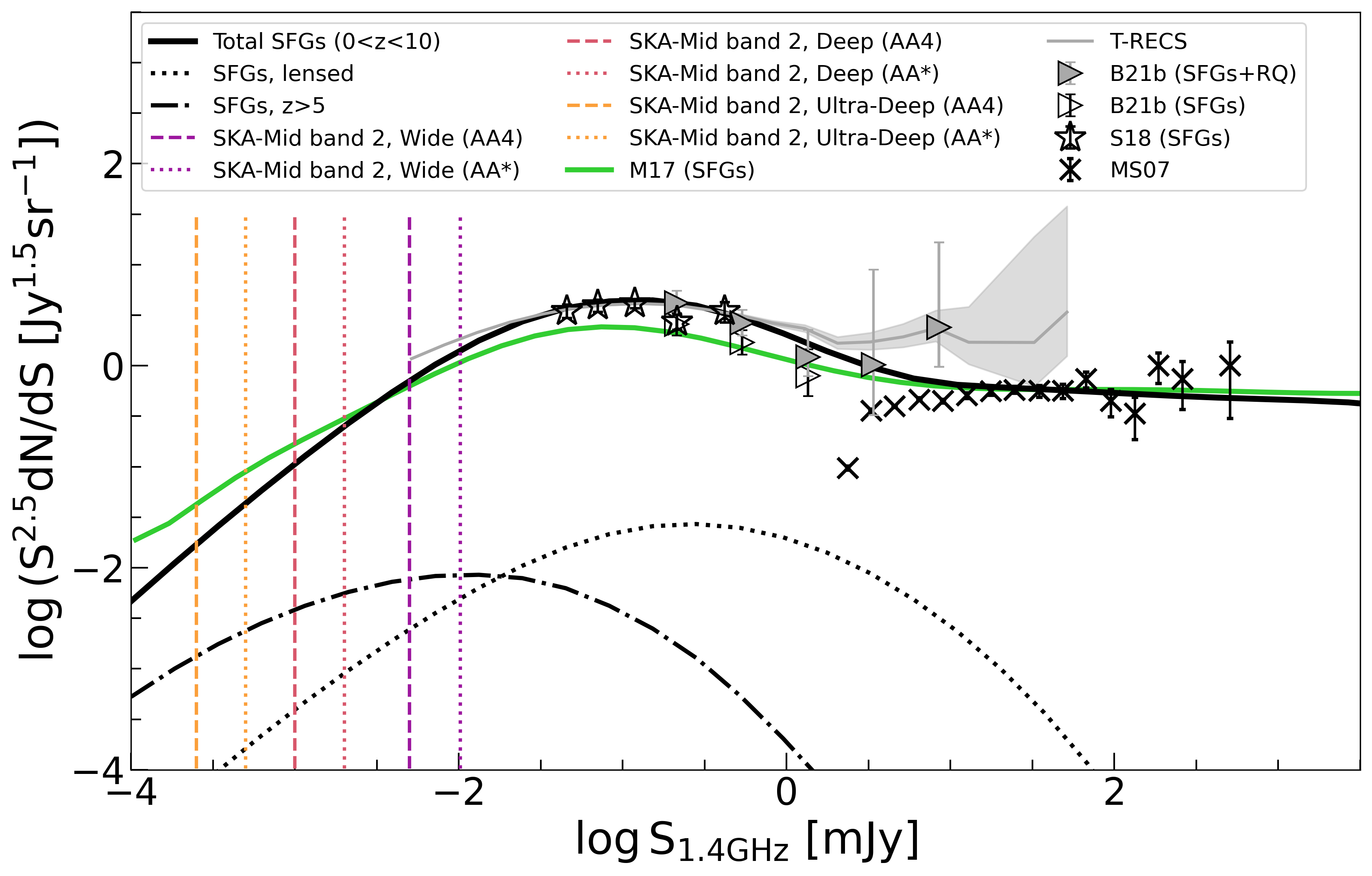}
    \caption{Euclidean normalised differential number counts from \texttt{SEMPER} for total (black solid line), high-z ($5\leq z\leq10$, dashdotted line), and lensed SFGs (black dotted line) at 1.4\,GHz. The solid green line is the prediction from \cite{Mancuso2017}. Triangles are data from \cite{Bonato2021a} for SFGs (empty) and SFGs and RQ sources combined (grey filled). Black crosses show the number counts for local galaxies from \cite{Mauch2007}. Stars mark the SFGs from \cite{Seymour2008}. The grey shaded area displays the predictions from the T-RECS simulation by \cite{Bonaldi2019MNRAS.482....2B, Bonaldi2023MNRAS.524..993B}. The dashed and dotted vertical lines represent the $5\sigma$ limits for the three SKA-Mid Band 2 surveys adopting the AA4 and AA$^*$ configuration, respectively.}
    \label{fig:number_counts}
\end{figure}

\cite{Mancuso2017} found that strongly lensed SFGs at high redshift contribute about 1\% to the radio number counts at $S_{\rm 1.4\,GHz}\approx 0.5\,$mJy. Therefore, we also compute the number counts for strongly lensed SFGs by accounting for the probability distribution of the magnification $d p/d\mu$ derived by \cite{Lapi2012}.  
The differential number counts for galaxy--galaxy lensing in the radio band are expressed as:

\begin{equation}
 \frac{d N^2_{\mathrm{lens}}}{d \log S_\nu d \Omega}\left(S_\nu\right)  =\int d z_s \frac{1}{\langle\mu\rangle} \int^{\mu_{\max }} d \mu \frac{d p}{d \mu} \frac{d^2N}{d \log S_\nu d \Omega}\left(S_\nu / \mu\right),
\end{equation}

where $z_s$ is the redshift of the lensed galaxy, $\mu$ is the magnification factor. Here, we assume $\mu_{\rm max} \approx 25$ for sources extended up to a few kiloparsecs, and $<\mu>$ is approximated as unity for the case of large-area surveys.
The resulting number counts are shown in Fig. \ref{fig:number_counts}.

\subsection{The Cosmic Star Formation Rate Density}

We can exploit \texttt{SEMPER} to make predictions on the evolution of the Cosmic SFRD. Given the SFRF, the SFRD ($\rho_{\psi}$) can be computed as:

\begin{equation}
    \rho_{\psi} (z) = \int d\log\psi\, \psi \frac{d^2N}{d\log\psi dV} (\psi,z).
\end{equation}

By substituting Eq. \ref{eq:sfrf}, one obtains:

\begin{equation}\label{eq:sfrd}
    \frac{d\rho}{d\log M_{\star}} (z|M_{\star}) = \left( \int d\log \psi\, \psi \frac{dp}{d\log\psi}  \right) \times \frac{d^2N_{\rm SFG}}{d\log M_{\star}dV}(z,\log M_{\star}).
\end{equation}

\section{Predictions for Square Kilometre Array's surveys}\label{sec:ska_pred}

\begin{figure}
    \centering
    \includegraphics[width=0.8\linewidth]{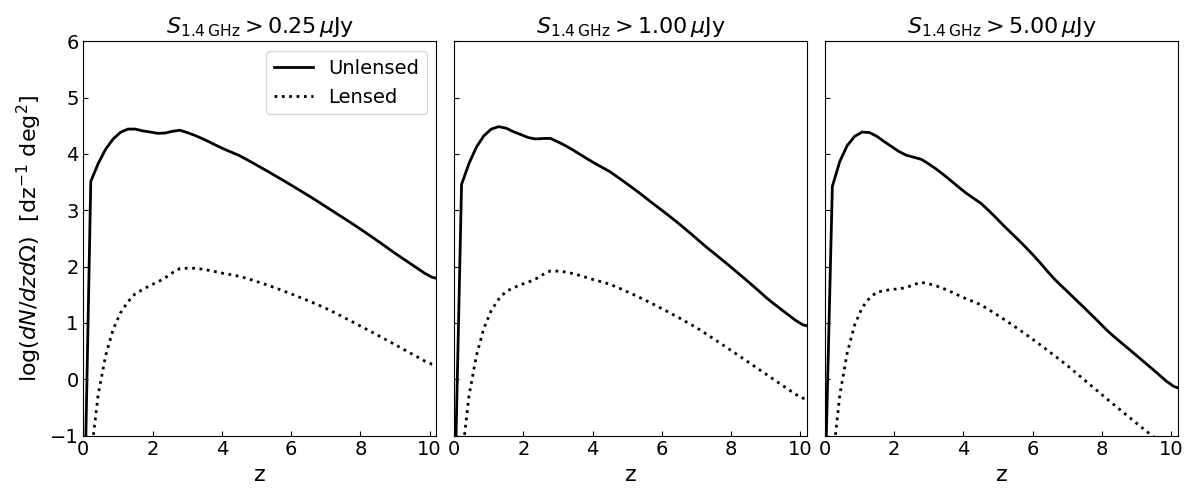}
    \caption{Predicted redshift distributions of the unlensed and lensed SFGs for the three SKA-Mid Band 2 surveys, computed adopting $5\sigma$ flux limits.}
    \label{fig:int_number_counts}
\end{figure}

The SKA-Mid AA4 configuration will consist of 197 dishes operating across five frequency bands. The range spanning approximately 1 and 1.4 GHz, corresponding to bands 1 and 2 of SKA-Mid, has been selected for several planned surveys \citep[][and see also \citealt{Prandoni01.2026.SKA} in this volume]{Prandoni2015}. Surveys with galaxy and AGN co-evolution as science driver are divided into Ultra-Deep, Deep, and Wide, covering areas of 1 deg$^2$, 10–30 deg$^2$, and about 10$^3$ deg$^2$. These surveys reach rms sensitivities of \SI{0.05}{\micro\jansky}, \SI{0.2}{\micro\jansky}, and \SI{1}{\micro\jansky} rms, respectively, corresponding to continuum integration times of approximately 2267, 167, and 6.7 hrs per pointing, adopting a Briggs weighting scheme for the AA4 configuration (see \citealt{Prandoni01.2026.SKA} in this volume for details). For the same rms sensitivities, but adopting the AA$^*$ configuration, the continuum integration times correspond to \num{11111}, 698, 28 hrs. Confusion limit affects only the Ultra-Deep survey, and a minimum angular resolution of 2 arcseconds is required for the AA$^*$ configuration to avoid it.
In Fig. \ref{fig:number_counts}, we report the radio number counts for unlensed and lensed SFGs at 1.4\,GHz predicted by our model, along with the $5\sigma$ limits for the three SKA-Mid surveys shown as dashed (dotted) vertical lines for the AA4 (AA$^*$) configuration. Our results are compared with observations of SFGs from \cite{Mauch2007}, \cite{Seymour2008}, and \cite{Bonato2021a}, with the T-RECS simulation (\citealt{Bonaldi2019MNRAS.482....2B, Bonaldi2023MNRAS.524..993B}), and with the semi-empirical model from \cite{Mancuso2017}. The SKA-Mid Band 2 Wide survey conducted with the AA$^*$ configuration will reach an rms sensitivity of \SI{2}{\micro\jansky}, enabling the detection of SFGs up to $z\approx10$ at $5\sigma$, corresponding to flux densities about 0.6\,dex fainter than those currently observed. With the more extended AA4 configuration, the survey will push the detection limit down to $\approx$1\,dex fainter than current observations for SFGs.
In particular, we predict at least $\approx 3\times10^4$ SFGs up to z$\approx10$ per square degree for the wide survey. For comparison, a deep radio survey at a similar frequency, such as the COSMOS-VLA 3\,GHz Large Project (see e.g. \citealt{Novak2017}) obtained with 384 hours of observations, found $\approx 6000$ SFGs up to $z\approx 5$ over 2\,deg$^2$ with flux densities higher than \SI{11.5}{\micro\jansky}. Moreover, we predict 5136, 1364, and 232 SFGs at $5 \le z \le 10$ at $S_{\mathrm{1.4\,GHz}} > \SIlist{0.25;1.0;5.0}{\micro\jansky}$, corresponding to the $5\sigma$ depth of the SKA-Mid Ultra-Deep, Deep, and Wide surveys, respectively (see Fig. \ref{fig:int_number_counts}). For the same depths, we predict 489, 275, and 124 strongly lensed SFGs. Moreover, we estimate that lensed sources contribute approximately 0.6\% to the total counts of SFGs.

\begin{figure}
    \centering
    \includegraphics[width=0.7\linewidth]{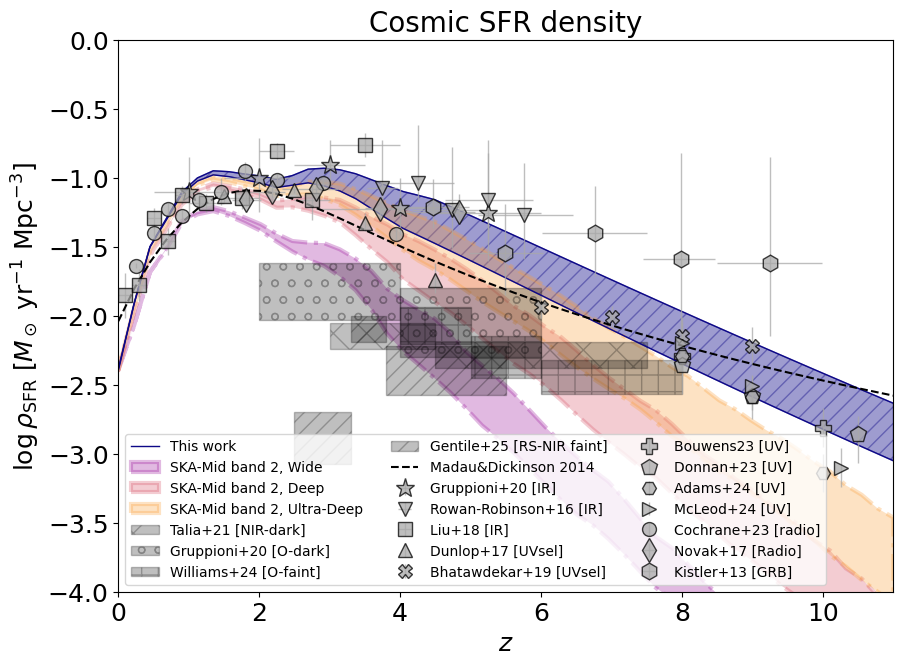}
    \caption{Predictions from \texttt{SEMPER} for the Cosmic SFRD compared with results from various multi-band surveys. The blue-hatched area represents the total contribution from SFGs as predicted by our model. The purple, red, and orange areas represent \texttt{SEMPER}’s predictions for the SKA-Mid Band 2 surveys at a $5\sigma$ flux limit. The lower limit of each area represents the minimum fiducial value of the SFRD from our model, while the scatter reflects the uncertainty in the starburst fraction at low masses. For each prediction, the width of the coloured area reflects the range of starburst fractions assumed in the model.
    Data are from \citet[][hexagons]{Kistler2013}, \citet[][inverse triangles]{Rowan-Robinson2016}, \citet[][triangles]{Dunlop2017}, \citet[][rhomboids]{Novak2017}, \citet[][squares]{Liu2018}, \citet[][crosses]{Bhatawdekar2019}, \citet[][stars]{Gruppioni2020}, \citet[][plus signs]{Bouwens2023}, \citet[][circles]{Cochrane2023}, \citet[][pentagons]{Donnan2023}, \citet[][rotated hexagons]{Adams2024}, and \citet[][rotated triangles]{McLeod2024}. Data referring to high-z dusty galaxies are from \citet[][grey shaded area hatched with circles; IR-selected HST-dark sources]{Gruppioni2020}, \citet[][grey shaded area hatched with crosses; radio-selected NIR-dark sources]{Talia2021}, \citet[][grey shaded area hatched with stars; radio-selected NIR-faint sources observed with JWST]{Gentile2025}, and \citet[][grey shaded area hatched with plus signs; O-dark sources observed with JWST]{Williams2024}. The dashed line is from \cite{Madau2014}.}
    \label{fig:sfrd}
\end{figure}

In Fig. \ref{fig:sfrd} we show the Cosmic SFRD predicted by our model, along with the contribution provided by SKA-Mid Band 2 at a $5\sigma$ flux limit for the Ultra-Deep, Deep, and Mid surveys. We compare our predictions with results from various multi-band surveys conducted at different wavelengths.

We estimate that, with less than 20 hours of SKA-Mid Band 2 observations using the AA4 configuration, we can detect at least 20\% of the total SFRD predicted by \texttt{SEMPER}, including the contribution from dark galaxies up to redshift $\approx6$ (see e.g. \citealt{Talia2021}, \citealt{Enia2022ApJ...927..204E}, \citealt{Gentile2024a, Gentile2025}). 
For reference, the selections presented by \cite{Talia2021} and \cite{Gentile2024a} identified about \num{300} NIR-dark galaxies within the COSMOS-VLA 3,GHz Large Project, out of a parent sample of $\gtrsim 6000$ SFGs over approximately 2 deg$^2$.
According to our predictions for the planned SKA-Mid Band 2 surveys, more than $\approx 1500$ dark sources per deg$^2$ are expected above \SI{5}{\micro\jansky} at 1.4\,GHz. This corresponds to a substantial increase in the number statistics compared to current deep radio surveys, enabling the construction of significantly larger and more representative samples of dust-obscured systems.
In addition to sensitivity, the wide sky coverage of the planned SKA surveys will strongly mitigate cosmic variance, which currently limits studies of rare, massive, dust-enshrouded galaxies.
Importantly, the SFRD fraction probed at these flux limits coincides with the regime where obscured star formation is expected to provide a significant contribution to the total cosmic SFR budget at high redshift. SKA observations will therefore allow a direct determination of the obscured component of the SFRD and offer the possibility to distinguish between different evolutionary scenarios for massive SFGs at $z > 3$.

\section{Conclusions}\label{sec:summmary}

We presented an updated version of the Semi-EMPirical model for Extragalactic Radio emission (\texttt{SEMPER}, \citealt{Giulietti2025}) and provided predictions for planned SKA surveys \citep{Prandoni2015}.
In particular, we computed the differential number counts at 1.4\,GHz for lensed and unlensed SFGs and the cosmic star formation rate density (SFRD) up to redshift $\approx 10$ for radio-bright SFGs.
SKA surveys will probe the faintest flux density regimes ($S_{1.4\,\rm GHz}<0.1$\,mJy), corresponding to a range dominated by galaxies whose radio emission arises primarily from star formation, and will account for a significant number of high-z ($5\leq z\leq 10$) SFGs ($\gtrsim 200$) and strongly lensed SFGs ($\gtrsim 100$). We also showed that at least $\approx20\%$ of the total SFRD predicted by \texttt{SEMPER} can be recovered with $\lesssim$20 hours of SKA-Mid Band 2 observations, including contributions from the observed optically/NIR-dark galaxies up to z$\approx$6. Our model serves as a bridge between existing deep surveys and the upcoming SKA era, offering predictions for the evolution of radio emission in SFGs throughout cosmic time, which will be tested with future observations.

\section*{Acknowledgements}

We acknowledge support from INAF under the following funding schemes: Large Grant 2022 (project "MeerKAT and LOFAR team up: a Unique Radio Window on Galaxy/AGN co-Evolution") and Large GO 2024 (project "MeerKAT and Euclid Team up: Exploring the galaxy-halo connection at cosmic noon"). Part of the research activities described in this paper were carried out with contribution of the Next Generation EU funds within the National Recovery and Resilience Plan (PNRR), Mission 4 - Education and Research, Component 2 - From Research to Business (M4C2), Investment Line 3.1 - Strengthening and creation of Research Infrastructures, Project IR0000034 – “STILES - Strengthening the Italian Leadership in ELT and SKA”.

\bibliographystyle{abbrvnat-maxbibnames4}
\bibliography{chapter}

\end{document}